\documentstyle[psfig]{mn2e}

\def\etal{{\rm et al. }}

\begin{document}

\title{Galaxy Pairs in the  2dFGRS II. Effects of interactions
on star formation in groups and clusters}

\author[Alonso et al.]{M. Sol Alonso $^{1,2}$,
Patricia B. Tissera  $^{1,3}$,
Georgina Coldwell $^{1,4}$ and
Diego G. Lambas $^{1,4}$\\
$^1$ Consejo Nacional de Investigaciones Cient\'{\i}ficas
y T\'ecnicas.\\
$^{2}$ Complejo Astron\'omico El Leoncito, Argentina \\
$^{3}$ Instituto de Astronom\'{\i}a
y F\'{\i}sica del Espacio, Argentina.\\
$^4$ Observatorio Astron\'omico
de la Universidad Nacional de C\'ordoba,  Argentina.\\
}

\date{\today}

\pagerange{\pageref{firstpage}--\pageref{lastpage}}

\maketitle

\label{firstpage}

\begin{abstract}

We analyse the effects of galaxy interactions on star formation
in groups and clusters of galaxies with virial masses in the range
$10^{13} - 10 ^{15}  M_{\odot}$.
We find a trend for galaxy-galaxy interactions to be less
efficient in triggering star formation in high density regions
in  comparison with galaxies with no close companion.
 However, we obtain the same
relative projected distance and relative radial velocity
thresholds
for the  triggering of  significant star formation activity
($r_p \sim 25 h^{-1} $ kpc and
$\Delta V \sim 100 \ {\rm km \  s^{-1}}$) as found in the field.
Thus, the nature of star formation driven by galaxy interactions is
 nearly independent of environment, although  there is a general lower
level  of star formation activity in massive systems.
The above results reflect, on one hand,  the local nature of star
formation induced by tidal interactions and, on the other,
the role played by the internal properties of galaxies.
By using a 2dFGRS mock catalog we estimate the contamination by spurious pairs
finding that  the statistics are cleary dominated by real pairs. We also
found  the behaviour of the trends to be robust against the use
of more restrictive relative
velocity thresholds.
We obtain a similar radial and relative velocity distribution of the pairs with respect
to the group centers compared to those of other typical group members, indicating that
galaxy pairs have no particular location and dynamics within groups.
We find that galaxy pairs in rich groups are systematically
redder and with a lower present-day star formation activity than
other group members.
A higher efficiency of galaxy-galaxy
interactions in dense regions in the past, or the fact that
early-type galaxies, the majority in groups, are
dominated by a central stellar spheroid being
more stable against tidal torques,
could provide suitable explanations to this observed trend.

\end{abstract}


\section{Introduction}

Different observations have shown that galaxy interactions are powerful mechanisms to trigger
star formation (e.g., Kennicutt et al 1998; Yee \& Ellingson 1995).
It has been also
found that both the rates of interactions and mergers
(Le Fevre et al. 2000; Patton et al. 2002)
and  the comoving star formation rate of
the Universe   increase
at least up to $z \approx 2$ (Stanway et al.  2003).
Barton et al. (2000) and Lambas et al. (2003) have carried out
statistical analysis of the star formation in pairs in the field finding
that proximity in radial velocity and projected distance  could
be linked to an increase of the star formation (SF) activity.
A similar conclusion is reached by Postman \& Geller (1984) from
the analysis of galaxies within and around voids.
By analyzing  1258 field galaxy
pairs  in the 100K 2dF public release,
Lambas et al. (2003, hereafter PaperI)
showed that galaxies with a close companion have significant
larger SF than  isolated galaxies with similar luminosity and
redshift distribution only if they are at a projected distance $r_{\rm p} <
25 h^{-1}$ kpc and  radial velocity $\Delta V < 100 \ {\rm km s^{-1}}$.
These authors also found that more luminous galaxies in pairs are the
most affected ones by interactions when compared to isolated galaxies with
the same luminosity, expect for similar luminosity pairs where both
members have enhanced SF.

Numerical studies of galaxy collisions
probe that gas inflows can be triggered during galaxy interactions producing
a sharp increase of the star formation activity even before the actual fusion
of the galaxies, depending on the orbital parameters and the internal
dynamical properties of the systems (e.g.,
Mihos 1992; Mihos \& Hernquist 1994, 1996; Barnes \& Hernquist 1996).
 It is well known that
galaxies dominated by a central stellar component are more stable
against tidal torques.
 In this case, tidal interactions are less efficient
in inducing star formation during the orbital decay phase
(Binney \& Tremaine 1987).
By using cosmological simulations, Tissera et al. (2002) found that
the response of a galactic system to tidal interactions varies along
it evolutionary history, expected to be stronger at
 early stages
of evolution when the systems have internal properties not suitable
for providing stability.

On the other hand, it is well known that galaxies in groups and clusters have
significantly reduced star formation  with respect to
those in the field and that, the star formation
activity depends on the distance to the centre
(Mart\'{\i}nez et al. 2002; Dom\'{\i}nguez et al.
2002). However, it is still uncertain how relevant  the global environment is
in the regulation of the star formation in galaxies.
Loveday, Tresse \& Maddox (1999) found that galaxies with prominent
emission-lines display weaker clustering than more quiescent galaxies.
Tegmark \& Bromley (1999) also found that early spectral types are
more strongly clustered than late spectral types. Besides,
Carter \etal (2001) suggest that the triggering of star formation
occurs on a smaller spatial scale and whether a galaxy forms stars
or not is strongly correlated with the surrounding galaxy density
averaged over a scale of a few Mpc.
Lewis \etal (2002) confirmed this last result by studying
 the environmental dependence of galaxy star-formation
rates near clusters, finding that
it is insensitive to the global
large-scale structure in which the galaxy is embedded.
 The authors also obtained that the distribution of
star-formation rates is correlated with both the distance from the cluster
centre and the local projected density (see also Dom\'{\i}nguez et al. 2002).

Taking into account these results,
in this work  we focus on the analysis of galaxy-galaxy interactions within
groups and clusters with the aim at assessing if this physical mechanism
plays a significant  role in star formation triggering.
For this purpose we constructed
the  hitherto  largest sample of interacting pairs in groups and
clusters  from the 2dFGRS.
By means of spectroscopy and colour analysis, we explore
the dependence of the star formation
in galaxy pairs on relative projected separation, radial velocity and
groupcentric distance.

\section{Data and Analysis}

\subsection{Pairs in groups in the 2dFGRS}

The 2dF Galaxy Redshift Survey comprises over 220000 spectra of
galaxies located in two contiguos declination strips (Colles et al 2001).
The spectral properties of 2dFGRS galaxies are characterised
using the principal component analysis (PCA) described
by Madgwick et al. (2002). This analysis makes use of the spectral
information in the rest-frame wavelength range 3700{\AA} to 6650{\AA}, thereby
including all the major optical diagnostic between OII and H$\alpha$ line.
For galaxies with $z > 0.15 $, the relation between the derived
star formation rates and the spectral classification can be
  affected by  sky absortion bands contamination of  the
H$\alpha$ line.
Consequently, we restrict to
galaxy pairs at $z\leq 0.1$ in order to prevent the results from strong
biases.
The 2dFGRS spectra are classified by a parameter, $\eta$, which is a linear
combination of the first and second principal components which
isolates the relative strength of
emission and absorption lines present in each galaxy spectrum.
Physically, $\eta$ is related to the specific star formation rate in a galaxy,
given the correlation with the equivalent width of H$\alpha$
found in emission lines galaxies (Bland-Hawthorn et al. 2002).
Galaxies with low star-formation rates
have typical values $\eta < -1.4$ and actively star forming systems
$\eta > -1.4$ (Madgwick \etal 2002).
We study star formation induced by tidal interactions
estimating the stellar birthrate parameter,  $b=SFR/<SFR>$
which  indicates the
present level of star formation activity of a galaxy related to its
mean past history.
Following Paper I, throughout this work  we
use the linear correlation between
$b$ and $\eta$, $b = 0.25  \eta + 1.06$,
as an estimate of the star formation activity in 2dFGRS galaxies.

In Paper I, we analysed a sample of 1853  galaxy  pairs in the 100 K
release of the 2dF galaxy redshift survey defined by
 a projected distance ($r_{\rm p} = 100 h^{-1} $ kpc
 and a relative radial velocity ($\Delta V =  350 {\rm km s^{-1}}$.
(We adopt in this paper $H_0= 100 h  {\rm km s^{-1} Mpc^{-1}}$). These limits
proved to be  reliable ones to select  interacting pairs with enhanced
 star formation activity.
By applying the same selection criteria, we  identified
a total of 9174  pairs in the 2dFGRS.
In order to analyse in detail the properties of galaxy
interactions in high density environments we constructed a
catalog of pairs in groups by cross-correlating
the total galaxy pairs catalog with the 2dFGRS group catalog
obtained by Merch\'an \& Zandivarez (2004, in preparation).
These authors identified
groups by  using a slightly modified version of the group
finding algorithm developed by Huchra \& Geller with a minimum number
of 4 members, an outer number density enhancement of 80 and
a linking radial cutoff of 200 ${\rm km s^{-1}}$.
The sample in the catalog comprises 6076 groups
spanning over the redshift range of $0.003 \le z \le 0.25$ with a mean
redshift $ z \simeq  0.1$. As a result of this
cross-correlation we obtain a sample of  4658
galaxies pairs in groups.

Despite of the fact that the 2dF public catalog is not complete,
we argue that galaxy  pairs searching is not
severely  affected by completeness effects.
This is based on the fact that although  the minimum fiber separation for
2dF spectroscopy is approximately 25 arcsec, the survey strategy was
to repeat the measurements in each field with new fiber positions
in order to achieve the highest completeness. Thus, from this point of view
there is no bias against small angular separations which would introduce spurious
results, specially at higher redshifts. Therefore, the inclusion of a pair
in our catalog depends mostly on the inclusion of each member in the survey,
which were randomly selected within the target of each field.
We argue that there are not significant selection effects on the pair sample which
could bias our statistical results on star formation activity.

Following the procedure outlined of Paper I, we focus our attention on the
effects of interactions on star formation by comparing with a suitable control
 sample which differs from the pair catalog only on the fact that galaxies in groups in
the latter have a close companion.
Using Monte Carlo algorithm we select for each galaxy pair, two other members of the 2DFGRS
group catalog.
Therefore, in this paper, the control sample corresponds to 9316 galaxies
in groups and clusters which do not have a companion within
$r_{\rm p} <100 h^{-1}$  kpc and $\Delta V <  350 {\rm km/s}$.
We stress the fact that this comparison sample of galaxies in
groups shares the same environment and has the same redshift distributions
than the sample of galaxy pairs in groups.

 \subsubsection{Testing the effects of spurious galaxy pairs}

The selection of galaxy pairs by using projected velocity
differences ($\Delta V$) and projected separation ($r_{\rm p}$)
has the drawback that spurious pairs can be included.
The use of cut-offs for both variables helps to diminsh the problem,
although they do not solve it complete.
In particular, the effects of spurious pairs are
expected to be stronger in high density regions (Mamon 1986; 1987).

In order to assess the effects of spurious pairs in our observational analysis,
we have used the 2dFGRS mock catalog constructed  by Merch\'an and Zandivarez
(2002) from a  gravitacional numerical simulation of the concordance
 $\Lambda$
cold dark matter universe ($\Omega_m=0.3, \Omega_{\gamma}=0.7$, H = $70 {\rm
km s^{-1} Mpc^{-3}}$
and  $\sigma_8=0.9$). The authors performed
 this simulations by using the HYDRA N-body code developed
 by Couchman, Thomas \& Pearce
(1995) with $128^3$ particles in a cubic comoving volume of 180 $h^{-1}$ Mpc
 per side, starting at $z=50$.

>From the 2dFGRS mock catalog a mock galaxy pair catalog was constructed by
applying the same observational cut-offs defined in PaperI: $r_{\rm p} < 100 \ h^{-1}$ kpc
and $\Delta V < 350 {\rm km s^{-1}}$ . Similary,
a close mock pair catalog was  obtained by requiring:  $r_{\rm p} < 25 \ h^{-1}$ kpc
and $\Delta V < 100 {\rm km s^{-1}}$.
The  orbital parameters such as major semi-axis, eccentricities, bounded
energy, etc.,  were estimated for both mock pair catalogs by assuming a two-body
problem scenario.
 We adopted $e <1$ and negative bounded energy
to distinguish between real  and spurious pairs.

We found that for the complete mock pair catalog, $71\%$ were real pairs while
for the close catalog, the percentage was larger: $79\%$.
We also  impossed the   condition of an extra neighbour
within  $r_{\rm p} < 400 \ h^{-1}$ kpc
and $\Delta V <  500 \ {\rm km s^{-1}}$ in order
 to segregate pairs according to environmemt. For galaxy pairs in
dense regions, we found that $73\%$ and $79\%$ were real pairs in
all and close pair mock catalogs,  respectively.
>From these estimations, we conclude that although spurious pairs
are present, real binary systems cleary dominate the statistics. In agreement with
previous work, we found that the contamination is larger in denser regions,
however, the differences with that of low density environment are not
significant, at least, when our  cut-off criteria are adopted.
Nevertheless, we have included estimations of the effects of spurious pairs
along the paper that can help to further assess their impact on the results.

\subsection{Position and velocity distribution of galaxy pairs in groups}

\begin{figure}
\centerline{\psfig{file=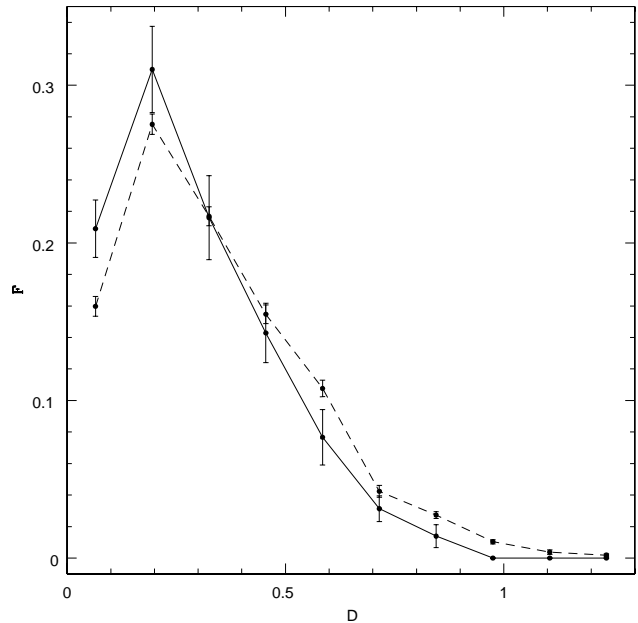,width=9cm,height=9cm}}
\caption{Distribution of normalised groupcentric distances $D$
for close galaxy pairs restricted to $r_{\rm p}< 25 h^{-1}$ kpc (solid lines)
and the control sample (dashed line).
}
\label{fig1}
\end{figure}

\begin{figure}
\centerline{\psfig{file=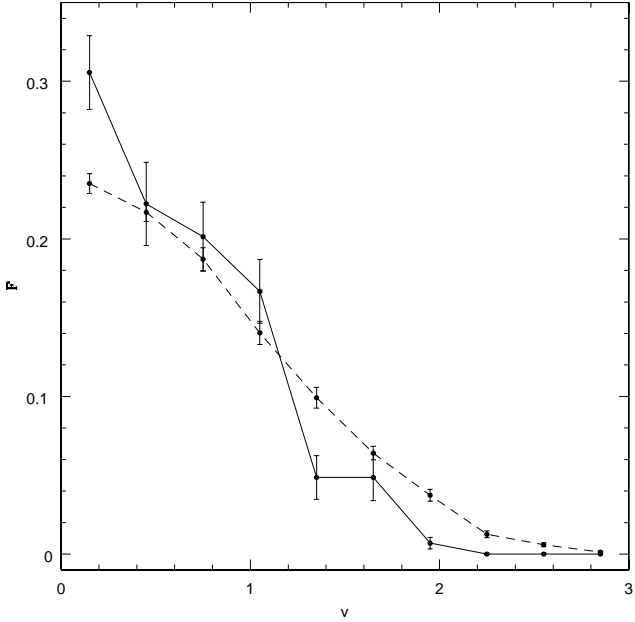,width=9cm,height=9cm}}
\caption{Distribution of normalised  radial velocities $v$
for close galaxy pairs restricted to $r_{\rm p}< 25 h^{-1}$ kpc (solid lines)
and the control sample (dashed line).
}
\label{fig2}
\end{figure}

\begin{figure}
\centerline{\psfig{file=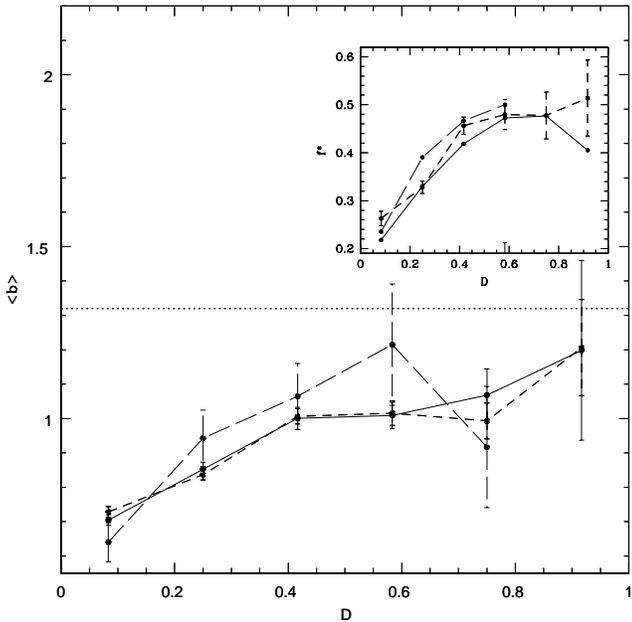,width=9cm,height=9cm}}
\caption{Mean $b$ parameter versus normalised groupcentric distance ($D$) for
galaxy pairs  (solid line), close pairs (long dashed line) and the control sample
(dashed line) in groups.
The dotted line corresponds to the mean $b$ value obtained in PaperI
for field galaxies.
The small box shows the fraction $f^\star$ of galaxies
with $b>\bar{b}$ in the galaxy pair (solid line), in close galaxy pairs
(long dashed line) and in the control sample (dashed line).
}
\label{brcentroran}
\end{figure}

We firstly investigate whether galaxy pairs have a particular radial location
in groups with respect to the control sample. For this purpose,
we have analysed the distribution of
projected radial distance, $R_D$, and relative velocity, $V$ ,
of pairs with respect to the host group centre
normalised to the group virial radius  ($R_{\rm Vir}$)
and group mean velocity dispersion ($\sigma$),
respectively ($D=R_D/R_{\rm Vir}$ and $v=V/\sigma$).
We restrict the analysis to  groups with more than 10 members
for the purpose of avoiding  large uncertainties in the determination of
 group centre, mean velocity and velocity dispersion
owing  to small number statistics. Since close pairs
are the most likely to have tidally enhanced star formation activity
we further restrict this analysis to close pairs $r_p < 25 \ h^{-1}$ kpc.
The resulting distributions of close pairs and of other member galaxies
are shown in Fig.\ref{fig1} and Fig. \ref{fig2},
from where it can be appreciated that close pairs and the other group members
are similarly concentrated and  have a comparable relative velocity
distribution  with respect to the host group centre.

\subsection{Comparison of the star formation in galaxy pairs and in the control sample}

We have also computed  the mean  star formation birthrate
parameter $b=SFR/<SFR>$
 for different bins of normalized groupcentric distance $D$
for the group galaxy pairs, the close pairs and the control sample.

The results from Fig.\ref{brcentroran} show clearly
that the star formation in galaxy pairs and in the control sample
strongly increases for larger groupcentric distance, approaching
the mean value for field galaxies in the outskirts.
The similarity of  these trends  in both samples shows, on average,
that the environment
has the same effects in all group members.
A similar behaviour is found for the close pairs althought
they  have larger star formation activity, as expected.

In order to improve our understanding of the star formation properties of
galaxy in pairs,
 we have  calculated the fraction of strong star forming
galaxies ($f^\star = N(b > \bar{b}_{\rm con})$), where
$\bar{b}_{\rm con}$ is the mean birth rate parameter of the corresponding
control sample (see also PaperI).
In the small box of Fig. ~\ref{brcentroran} we show the corresponding
$f^\star$ for the galaxy pair, the close pair  and the control samples  as a function
of the galactrocentric distance. This figure agrees with the trend
found for  the mean $b$,
 indicating that in the central regions the star formation
activity is weaker for pair, close pairs and control galaxies.

\subsection{Star Formation as a function of orbital parameters}

We analyse the dependence of star formation on pair relative
projected separation $r_p$
and relative  radial velocity  $\Delta {\rm V}$
estimating mean values $<b>$ as a function of $r_{\rm p}$
and $\Delta {\rm V}$ for our sample of galaxy pairs in groups.

\begin{figure}
\centerline{\psfig{file=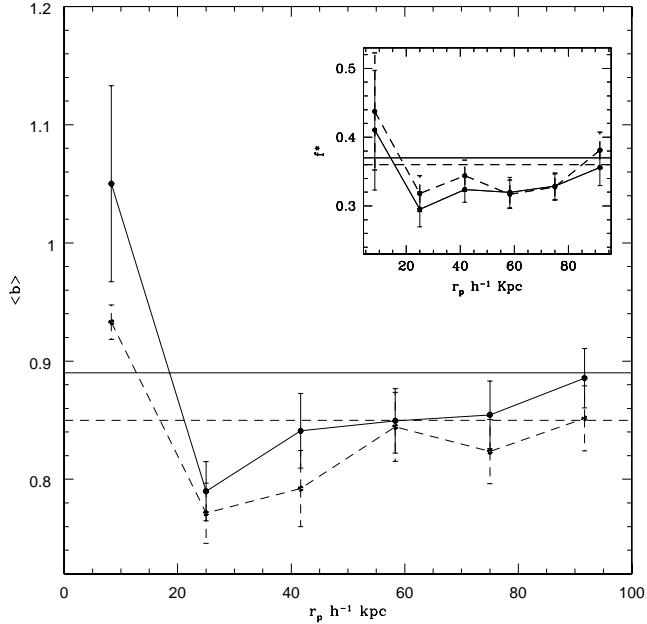,width=9cm,height=9cm}}
\caption{Mean $b$ parameter versus
projected distance for galaxy pairs in groups (solid line)
and pairs restricted to $D < 0.5$ (dashed line).
The small box correspond to the fraction $f^\star$ of galaxies
with $b>\bar{b}$.
}
\label{betarg}
\end{figure}

\begin{figure}
\centerline{\psfig{file=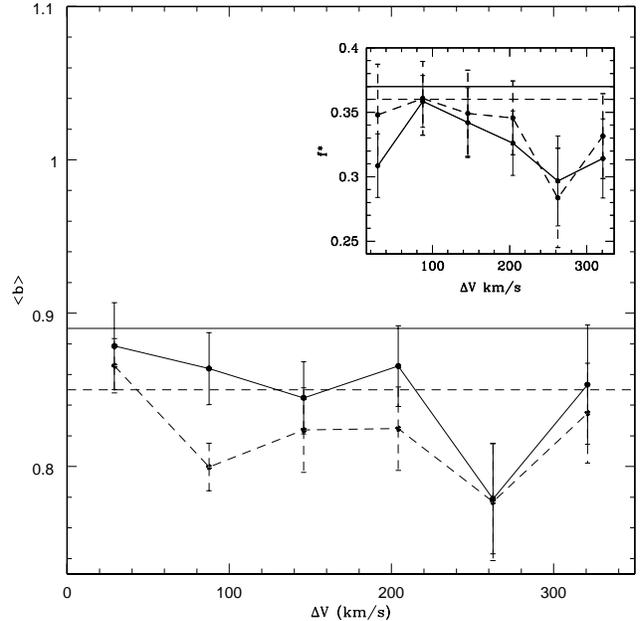,width=9cm,height=9cm}}
\caption{Mean $b$ parameter versus
relative velocity for galaxies in interacting pairs in groups (solid line)
and galaxies with $D < 0.5$ (dashed line).
The small box shows the fraction $f^\star$ of strong star forming galaxies
in the sample.
}
\label{betavg}
\end{figure}

The results are shown in Fig. \ref{betarg}
and Fig. \ref{betavg}. It can be seen that,
as it occurs in pairs in the field (Paper I), at $r_p< 25 h^{-1} $ kpc
the star formation activity is significantly enhanced over the control
sample.
Similarly, pairs with smaller relative radial velocity differences
have larger mean b values, consistent with the results found for field pairs.
This behaviour indicates that the physics of
star formation induced by pair interactions
operates in a similar fashion in high density environments,
although with a lower general level of star formation activity.

Given the reduced star formation activity of pairs in the central regions
of groups (Fig. ~\ref{brcentroran}),
 we have carried out a similar analysis adopting the restriction
$R_D/R_{Vir}<  0.5$.
The results for this subsample of centrally located pairs
are shown  in Fig. \ref{betarg}
and Fig. \ref{betavg} (dashed lines) from which it
can be seen the similar behaviour of the centrally located
pairs, albeit with an overall weaker star formation activity.

We detect a trend for the mean star formation birth rate parameter $b$
of close galaxy pairs to be more densely populated by early type galaxies
with  respect to the
corresponding mean  values ($\bar{b}$) of  the control samples
from  pairs in the field ($ <b>/\bar{b}_{\rm field} = 1.43 \pm 0.14$) to  pairs in the
densest regions ($<b>/\bar{b}_{\rm groups}= 1.09 \pm 0.01$). This trend indicates
that tidal torques generated by  interactions could be
less efficient in pair systems (with similar
orbital parameters) in  high density regions.
The fact that densest regions are more populated by early type
galaxies   suggests that the different response could due to
 differences in the dynamics of the galaxies in  pairs  and the
available gas reservoir
to form stars in these systems.

\begin{figure}
\centerline{\psfig{file=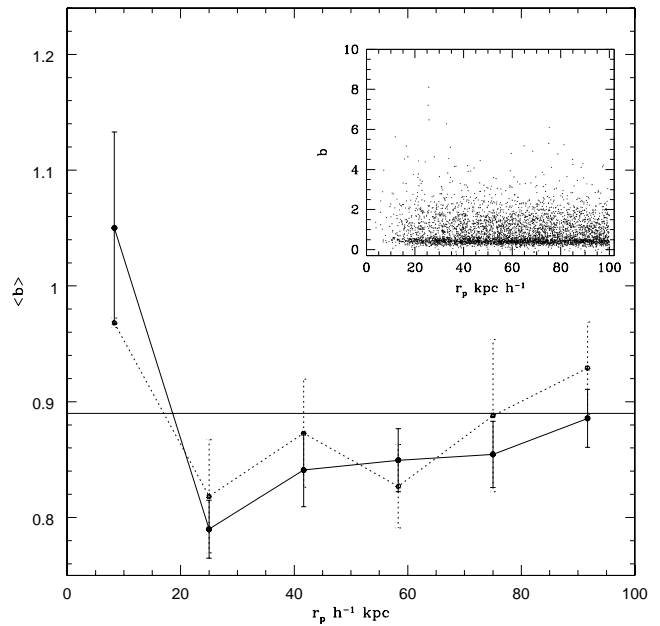,width=9cm,height=9cm}}
\caption{Mean $b$ parameter versus
projected distance for galaxy pairs in groups with
$\Delta V < 350 \ {\rm km s^{-1}}$ (solid line)
and pairs restricted to $\Delta V < 100 \ {\rm km s^{-1}}$ (dotted line).
The small box shows the points distribution of $b$ parameter versus
projected distance for pairs in groups.
}
\label{brvel}
\end{figure}

We should also take into account the possibility that spurious
galaxy pairs which are more probable in denser regions, can
contribute to diminish the signal, producing the observed trend.
In order to assess the effects of spurious pairs, we
use stricter difference velocity cut-offs, similarly
to the analysis we carried out in PaperI.
As we can see from Fig. \ref{brvel}, as pairs
with larger velocity differences are excluded, the mean b values
behave similar to close pairs. From this figure,
we see that spurious pairs could contaminate the results by
reducing the star formation activity signal but, their effects
are not significant enough to change the trends.
The small box in Fig.\ref{brvel} shows the points
distribution of $b$ parameter as a functionof  projected distance
for all galaxies in  pairs.
>From this figure it is clear that at small projected  separations there are few
points with low star formations activity.
Hence, the enhacement of the SF activity for very close pairs
seems not to be significantly biased by interpolers which can
artificially diminish it.

>From this analysis, we find no indications for an environmental dependence of the
relative separations and relative velocity thresholds for
 star formation induced by tidal torques. However,
our results might  indicate  a decrease in the efficiency of tidal torques
to induce SF as we move to denser environments (see also Section 2.6).

\subsection {Pair $b_j-R$ colours in groups }

Star formation in galaxies
can also be studied by analysing integrated colours which
reflect the relative fraction of old and new stellar populations.
We acknowledge that metallicity could introduce complications in this direct
interpretation, therefore this analysis is complementary of the previous one,
 serving  to deepen our understanding on star formation and galaxy evolution.
By providing an estimate of the fraction of new stars, galaxy colours
reflect a different timescale than emission lines. For instance, in
the case of an ongoing burst of star formation prominent emission lines
are expected although the integrated colour is not likely to be
affected.

\begin{figure}
\centerline{\psfig{file=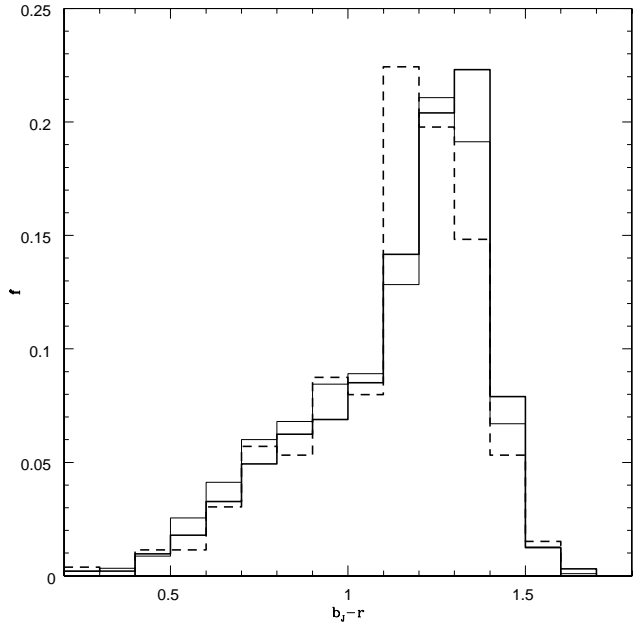,width=9cm,height=9cm}}
\caption{$b_j-R$ colour distributions of galaxies in pairs (solid thick line)
and in control sample (solid thin lines). We have also included the distribution
for the subsample of close  galaxy pairs:
$r_p < 25  $ kpc and $\Delta V < 100 \ {\rm km s^{-1}}$ (dashed thick line).
}
\label{hist1}
\end{figure}

Therefore, one may ask whether the distribution of colour indexes
of galaxies in pairs at a given spectral type differs from that
of  galaxies in the control sample.
For this purpose, we have analysed the distribution of $b_j-R$
colour indexes available in the 2dFGRS data release.
In Fig.\ref{hist1} we show the histograms corresponding to
$b_j-R$ colour distributions of pairs, close pairs and control sample galaxies
It can be appreciated in this figure
that the complete sample of galaxies pairs and galaxies in the control
sample have a remarkably similar distribution
of $b_j-R$ colour indexes, but close pairs galaxies restricted to
$r_p < 25 \ {\rm kpc}$ and $\Delta V < 100  \ {\rm km/s}$ have a distribution
shifted to the blue.

\begin{figure}
\centerline{\psfig{file=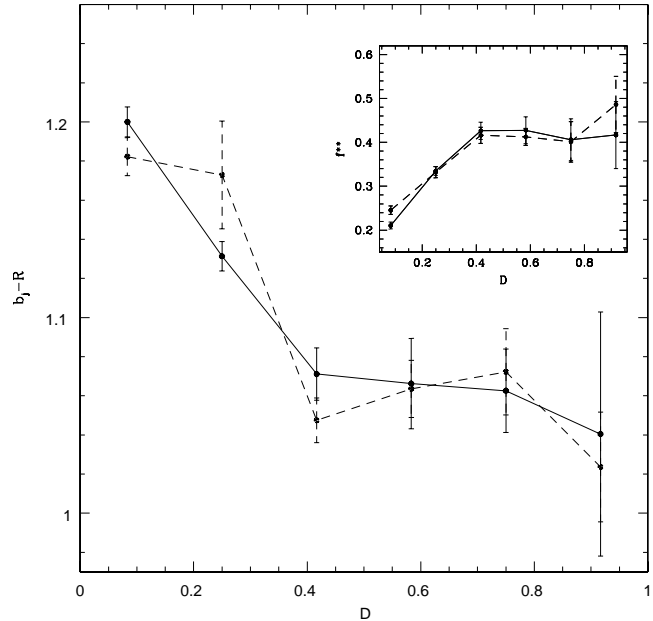,width=9cm,height=9cm}}
\caption{$b_j-R$ colour versus distance to the group centre
for   galaxy  pairs (solid line) and the control
sample (dashed line).
The small box shows to the fraction $f^{\star \star}$ of galaxies
 $b_j-R<\bar{b_j-R}$
}
\label{colD}
\end{figure}

In Fig. ~\ref{colD} we display the mean $b_j-R$ colours in bins of
groupcentric distance for galaxy pairs in groups and for the
corresponding the control sample.
This plot shows that galaxy pairs in groups have similar colours
than those without a close companion independently of the distance to the
group centre. This result is in agreement with  that shown in Fig. ~\ref{brcentroran}
where it was found a similar behavior for the mean birth rate parameter $<b>$.

Similarly to the estimation of the fraction of actively star forming galaxies,
we have calculated the fraction
$f^{\star \star} $ of galaxy in pairs with  $b_j-R < \bar{b_j-R}_{\rm con}$
where $\bar{b_j-R}_{\rm con}$ is the mean colour of the corresponding control sample.
In the small box of Fig. ~\ref{colD} we show the
$f^{\star \star}$ for galaxies in pairs and the corresponding control  sample
as a function of the galactrocentric distance.

\subsection{Results restricted to rich and poor group subsamples}

Important observational  evidence shows that dense environments
can affect many
galaxy properties, in particular the star formation rate (Dressler et. al 1985).
In studying this point, Dom\'{\i}nguez et al. (2002) found a clear
distinction between high and low virial mass groups (adopting
a mass limit of $10^{13.5} M_{\odot}$)
in the 2dF release 100 K  Group Galaxy Catalog constructed by
Merch\'an \& Zandivarez (2002).
 These authors
 found that  massive groups have a significant dependence of the
relative fraction of low star formations galaxies on local galaxy density
and groupcentric radius while low virial mass  groups
 show no significant trends.

In this subsection, and in order to have a sufficiently large number of pairs,
we have modified the 10 minimum number member condition to 4 members.
These enlarged  group sample is divided into
a poor and rich group subsamples at  $M_{\rm Vir} = 10^{13.5} M_{\odot}$.

\begin{figure}
\centerline{\psfig{file=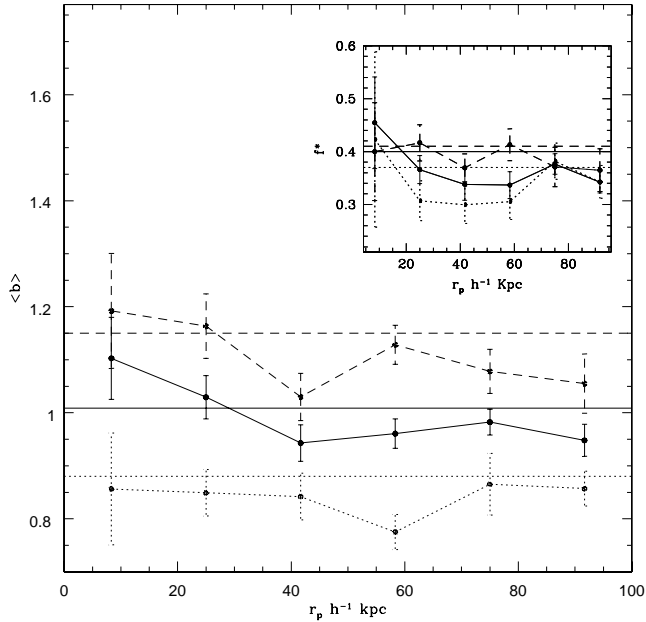,width=9cm,height=9cm}}
\caption{Mean $b$ parameter versus projected distance  for
galaxy pairs in all groups (solid line) and in the rich  and poor
subsamples (dotted and dashed lines, respectively).
The small box shows  the fraction $f^\star$ of galaxies
with $b>\bar{b}$ in the different subsamples (same code line).
Fractions have been estimated with respect to the corresponding
  $\bar{b}$ value of each subsample
}
\label{brcricos}
\end{figure}

\begin{figure}
\centerline{\psfig{file=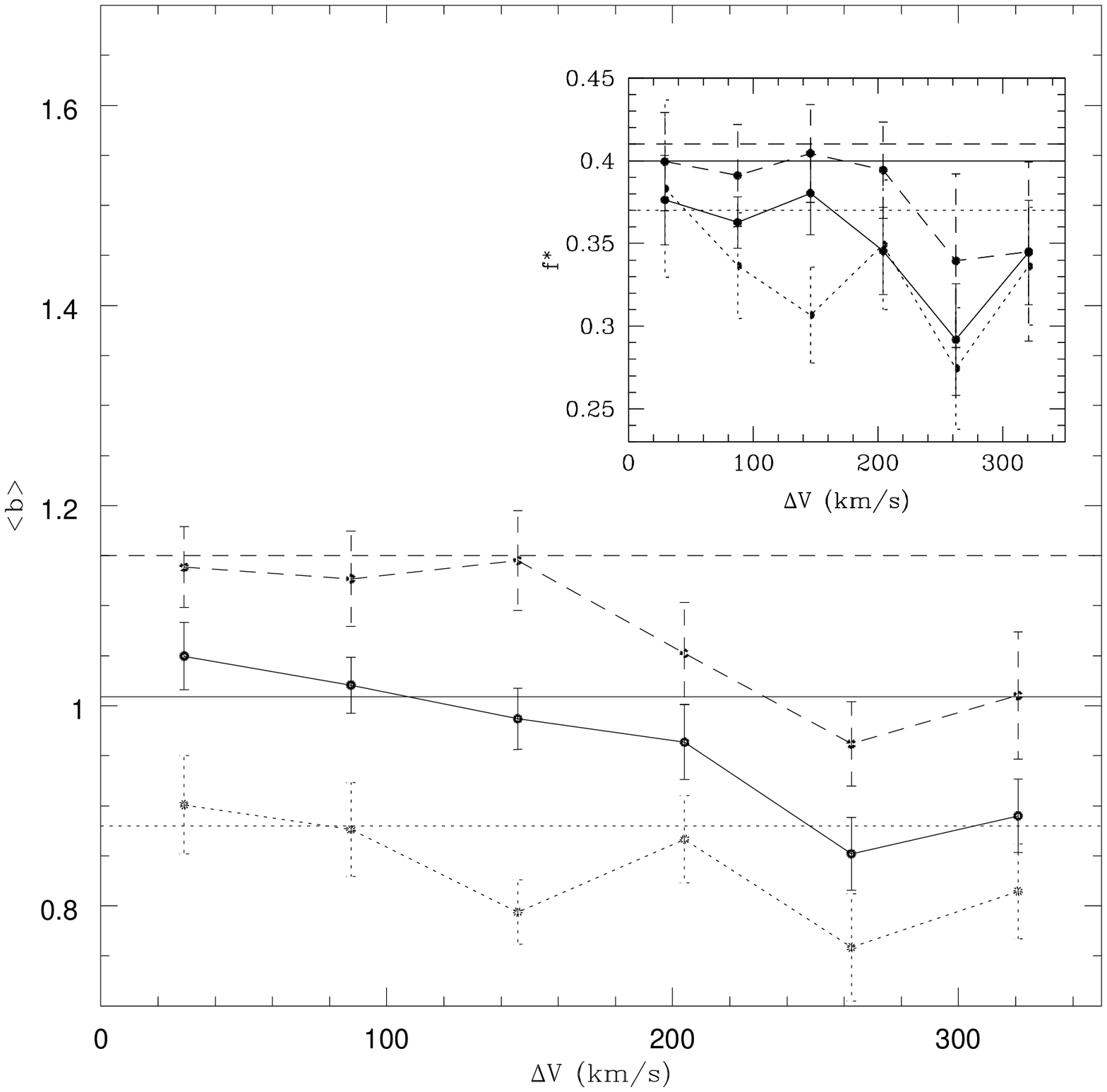,width=9cm,height=9cm}}
\caption{Mean $b$ parameter versus relative velocity  for
galaxy pairs in all groups (solid line)
and in the rich  and poor
subsamples (dotted and dashed lines, respectively).
The small box shows the fraction $f^\star$ of galaxies
with $b>\bar{b}$ in the different subsamples (same code line).
Fractions have been estimated with respect to the corresponding
  $\bar{b}$ value of each subsample.
}
\label{bvcricos}
\end{figure}

In Fig. \ref{brcricos} we show that, as expected, galaxy pairs in rich
groups show a weaker star formation activity  than galaxy pairs in the
total sample. We see that, in high mass groups,
there is no dependence of the  mean $b$ parameter on $r_p$.
Conversely, galaxy pairs in poor groups
have a higher star formation activity than
pairs in the total sample and show a significant dependence
of mean $b$ parameter on  $r_p$.
A similar behaviour appears for $b$ vs $\Delta V$  as
shown Fig. \ref{bvcricos}.
 This weak  dependence of $b$ on orbital parameters
in the high mass group subsample may be associated to a high efficiency of
star formation in these regions  in the past which
leaves little fuel for new bursts. Also, taking into account
the correlation between the spectral type $\eta$ and morphology
showed  by   Madgwick et al. (2002), the central regions of clusters
 would be mostly populated
by early type galaxies  dominated by a central stellar spheroid
 which can provide stability to the system against tidal torques
(e.g., Binney \& Tremaine 1987; Mihos \& Hernquist 1996; Tissera et al. 2002).
Hence, in these systems tidal field could  not be that effective to  drive
gas inflows and trigger starbursts even if there were gas available.

\begin{figure}
\centerline{\psfig{file=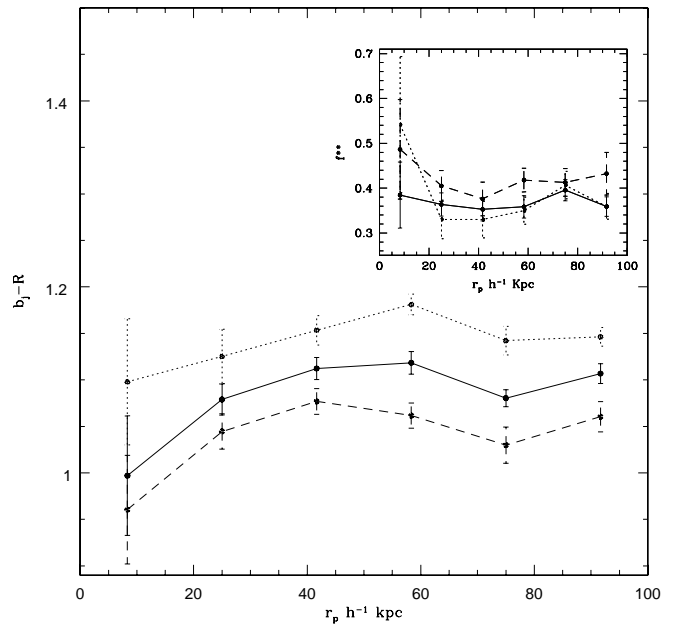,width=9cm,height=9cm}}
\caption{Mean $b_j-R$ galaxy colours versus projected distance for
galaxy pairs in all groups (solid line)
and in the rich  and poor
subsamples (dotted and dashed lines, respectively).
The small box shows the fraction $f^{\star \star}$ of galaxies
with $b_j-R<\bar{b_j-R}$ in all subsamples.
Fractions have been estimated with respect to the corresponding
  $\bar{b_j-R}$ of each subsample.
}
\label{colrgrupos}
\end{figure}

\begin{figure}
\centerline{\psfig{file=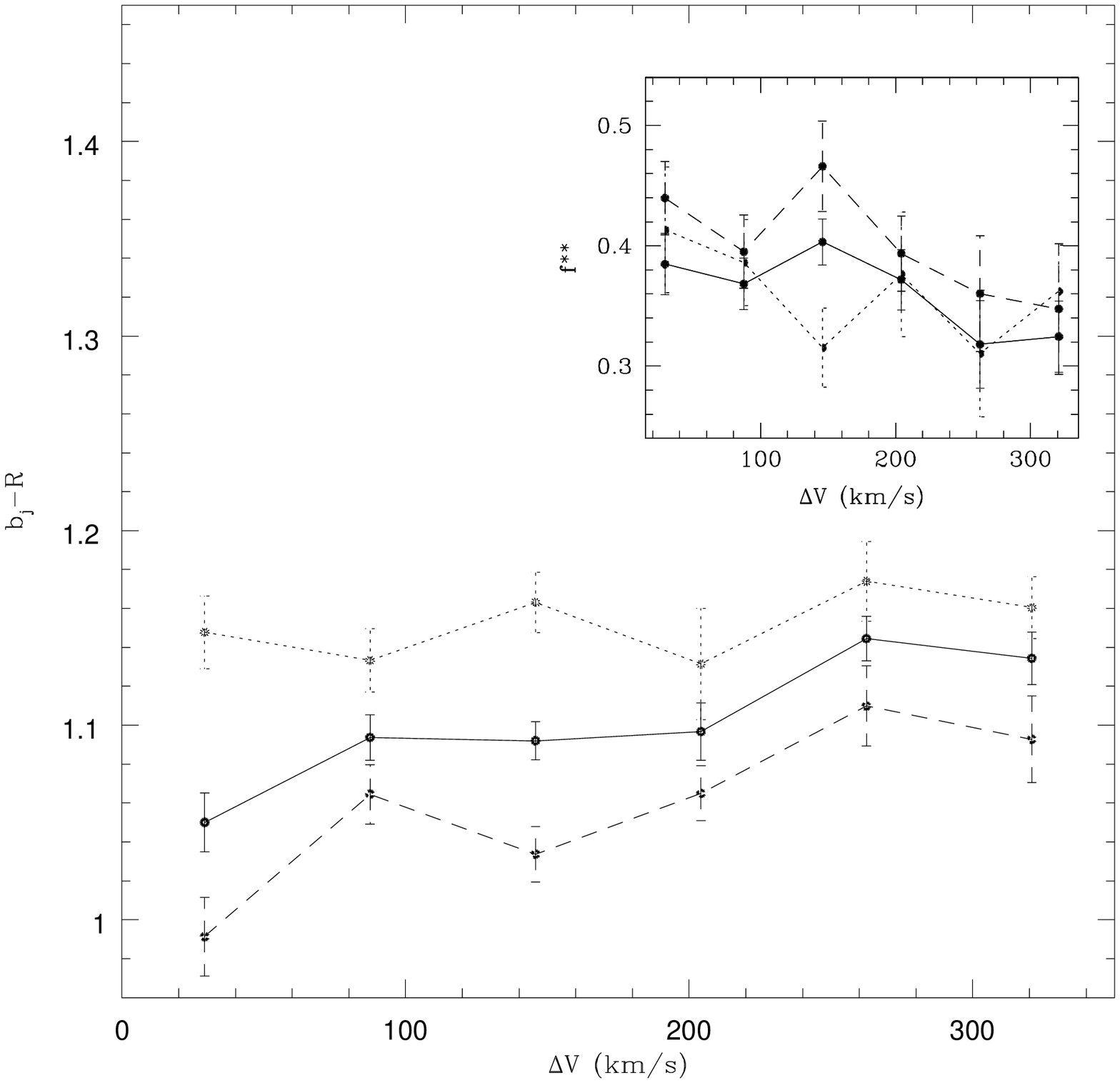,width=9cm,height=9cm}}
\caption{Mean $b_j-R$ galaxy colours versus relative velocity for
galaxy pairs in all groups (solid line)
and in the rich  and poor
subsamples (dotted and dashed lines, respectively).
The small box shows  the fraction $f^{\star \star}$ of galaxies
with $b_j-R<\bar{b_j-R}$ in the different subsamples.
Fractions have been estimated with respect to the corresponding
  $\bar{b_j-R}$ of each subsample.
}
\label{colvgrupos}
\end{figure}

In Fig \ref{colrgrupos}, we show that mean $b_j-R$  colours
of pairs in the rich group subsample are redder than those in the total
group sample.
It can be also seen a significant decline of the mean $b_j-R$ colours
at small projected separations ($r_p < 25 h^{-1}$ kpc)
for the three samples.
In Fig. \ref{colvgrupos} it can be appreciated
 a similar behaviour  for the mean $b_j-R$ colours as
a function of pair relative velocity $\Delta V$.

\begin{figure}
\centerline{\psfig{file=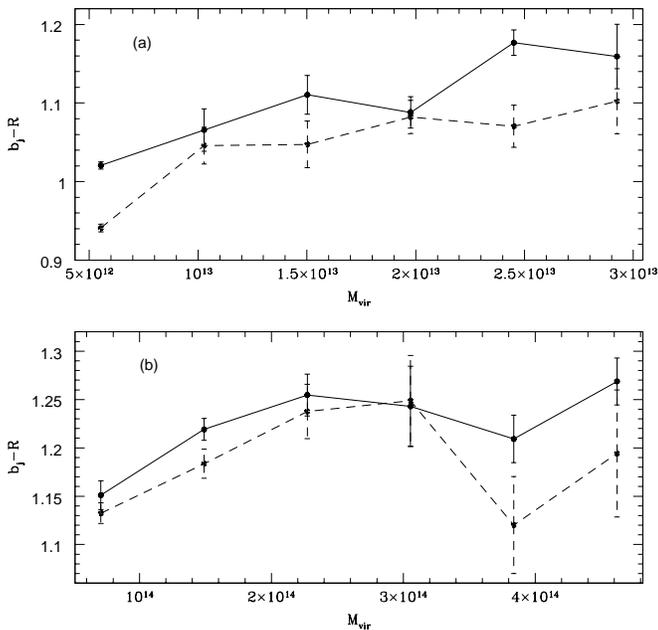,width=9cm,height=9cm}}
\caption{ Mean $b_j-R$ galaxy colours versus group virial masses for
galaxy pairs (solid line) and  control sample
(dashed line) in groups of low and high mass (a and b respectively).
}
\label{colmv}
\end{figure}

\begin{figure}
\centerline{\psfig{file=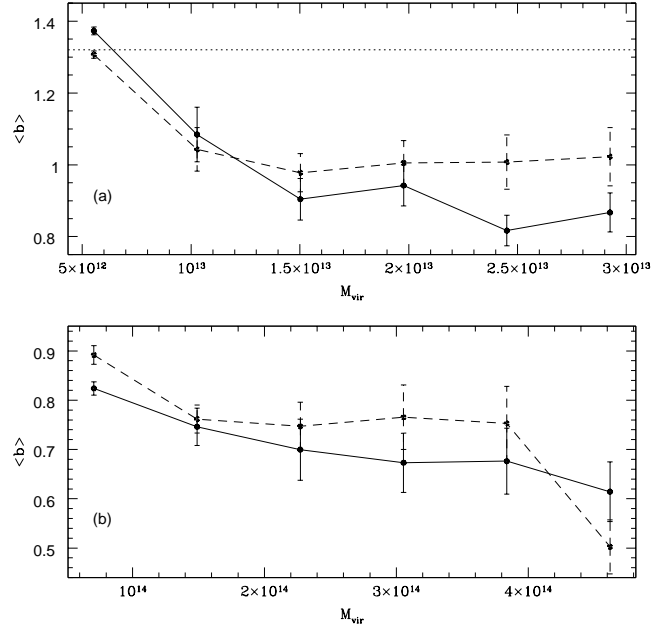,width=9cm,height=9cm}}
\caption{ Mean $b$ parameter versus group virial masses for
galaxy pairs (solid line) and  control sample
(dashed line) in groups of low and high mass (a and b respectively).
The dotted line represents the mean $b$ parameter of isolated
 galaxies in the field
(paper I).
}
\label{bmv}
\end{figure}

The $b$ parameters derived from the spectral type index $\eta$ have no resolution
for low star forming galaxies (Madgwick et al. 2002),
the majority in high mass groups. Thus, we computed the
fractions of star forming galaxies with $b> \bar{b}$ which shows similar
trends for rich and poor group sub-samples. Fig. ~\ref{colrgrupos} shows
that closer pairs are systematically red in both sub-samples, if
interlopers were a major problem we would not see the decreasing
trends with a projected separations ($r_p$).
Fig. \ref{colmv} shows the relation between mean $b_j-R$ colours for galaxies
in pairs and the host  group virial mass. Also we display the results for the
control sample.
We can see here that galaxy pairs are systematically redder
than those of the control sample.
These trends agree with those shown in
Fig. \ref{bmv}, from where  it can be concluded that
galaxies in pairs in rich groups have a present day lower star formation activity than
galaxies without close companions. Although effects by spurious pairs
should be always taken into account,
we argue this result is robust against spurious pairs  on the basis
of previous discussions (Sections 2..1.1 and 2.4)
and the fact that a further restriction to the difference velocity
cut-off yields similar relations albeit blueshifted.

The results of confronting the analysis of the pair and control catalogs,
suggest that  galaxy interactions  in rich groups
were more  efficient in triggering star formation so that
currently, these galaxies show less star formation activity
and have redder colours than other group members.

\section{Conclusions}

We have analysed the properties of pairs of galaxies
in high density environments
corresponding to groups and clusters of galaxies with virial masses
$10^{13} - 10 ^{15} M_{\odot}$. We stress the fact that the analysis discussed
in this paper is based on the comparison between two galaxy samples which differed between each
other {\it only} on the fact that one of sample comprises galaxies with  close companions,
and the other, do not.
Hence, we are always  estimating the statistical differences in
the properties of  galaxies  introduced
by the presence of a companion.

The main conclusions  can be summarized as follow:

\begin{itemize}
\item We find that the radial and relative velocity distributions
of close pairs, normalised to the group virial radius and group mean
velocity dispersion, with respect to the group centers
are similar to those of  group members with no companions indicating a no
particular location for galaxy pairs within the groups.

\item The general lower efficiency
of star formation in dense environments is accompanied by
a lower
 enhancement of the star formation induced by
interactions associated to galaxies of early-morphology
which  mostly populate rich systems.

\item  Similar relative projected separation and relative velocity
thresholds $r_p \sim 25 h^{-1}$ kpc and
$\Delta V \sim 100 \ {\rm km s^{-1}}$  within which
star formation is significantly enhanced with respect
to galaxies without a close companion,
is obtained for pairs in the field and in modest mass groups.

\item We find that star formation strongly increases for larger groupcentric
distance for both pairs and the control sample in a similar fashion.

\item The distribution of $b_j-R$ colour indexes of galaxies in close pairs
are bluer than of galaxies
in groups with no close companions.

\item For the subsample of pairs in high mass groups, we find a weaker dependence
of the star formation birth parameter $b$ on  relative projected
distance and relative velocity.
In these massive systems, galaxies in
pairs are systematically redder and with a lower present-day
star formation activity than galaxies in the control sample.

Galaxies in groups, dominated by a central stellar spheroid
would be more stable against tidal torques generated by interactions
(Binney \& Tremaine 1987; Mihos \& Hernquist 1996; Tissera et al. 2002)
which  would  not be very  effective in driving
gas inflows and triggering starbursts, even if there were sufficient gas available.
Also, we may consider the possibility that galaxy
interactions have been very efficient in stimulating star formation in the past
leaving little fuel for new bursts.

The analysis of a mock catalog of galaxy pairs derived from numerical simulationns
to mimic the
2dFGRS catalog showed that the statistics are not dominated
by spurious pairs. This finding is supported by the fact that the results are robust against
stricter velocity difference cut-offs.

\end{itemize}

\section{Acknowledgments}

We acknowledge useful discussions with Manuel Merch\'an and Michael Balogh.
We thanks the anonymous Referee for a detail  revision that
helped to improve this paper.
This work was partially supported by the
Consejo Nacional de Investigaciones Cient\'{\i}ficas y T\'ecnicas,
Agencia de Promoci\'on de Ciencia y Tecnolog\'{\i}a,  Fundaci\'on Antorchas
and Secretar\'{\i}a de Ciencia y
T\'ecnica de la Universidad Nacional de C\'ordoba.

\label{lastpage}
\end{document}